\documentclass[11pt]{article}

\usepackage{amsfonts,latexsym,eucal,color,graphicx,epsfig,amssymb,amsmath}

\newcommand{\be}{\begin{eqnarray}}
\newcommand{\ee}{\end{eqnarray}}
\newcommand{\beq}{\begin{eqnarray}}
\newcommand{\eeq}{\end{eqnarray}}

\newcommand{\dalm}{\kern1pt\vbox{\hrule height 0.9pt\hbox{\vrule width 0.9pt\hskip 2.5pt\vbox{\vskip 5.5pt}\hskip 3pt\vrule width 0.3pt}\hrule height 0.3pt}\kern1pt}

\usepackage{mathrsfs}

\begin{document}


\thispagestyle{empty}
\begin{center}

{\Huge \bf The Parts of the Gravitational Field}

\vspace{1.5cm}

{\large \bf Joan Camps}

\vspace{.5cm}
{\it Department of Physics and Astronomy, University College London\\
Gower Street, London WC1E 6BT, United Kingdom}\\
\vspace{.5cm}
\verb"j.camps@ucl.ac.uk"

\vspace{2cm}
24 May 2019

\vspace{2 cm}
{\bf Abstract}

\end{center}
It is argued that, unlike for other fields, the quasilocal parts of the gravitational field are always bounded by surfaces of extremal area.

\vspace{2cm}

\begin{center}
{\large \it Essay written for the Gravity Research Foundation 2019 Awards for Essays on Gravitation.}
\end{center}

\clearpage
\setcounter{page}{1}

The entanglement entropy of quantum fields across regions is UV divergent. This divergence is interesting, as it scales with the area of the entangling surface, suggesting a connection to gravitational entropy \cite{Bombelli:1986rw, Srednicki:1993im}.

This connection has a concrete realisation in AdS/CFT: the Ryu-Takayanagi formula, that equates entanglement entropy in the boundary quantum field theory with the area of certain surfaces in the gravitational bulk dual \cite{Ryu:2006bv, Hubeny:2007xt}. These bulk surfaces $\sigma$ have extremal area and are anchored at the field theory entangling surface -- in the conformal boundary of asymptotically AdS spacetimes.

It has been argued that the first quantum correction to the Ryu-Takayanagi formula is given by the entanglement entropy of the bulk fields across $\sigma$ \cite{Faulkner:2013ana}:
\beq
S=\frac{\left.\textrm{Area}\right|_{\sigma}}{4G}+\left.S_{\textrm{bulk}}\right|_{\sigma}+\cdots.
\label{eq:Sgen}\eeq
The area-divergence of the second term is absorbed by a renormalisation of Newton's constant, and thus the rhs of this expression is UV finite.

The first instance of a formula of the type of \eqref{eq:Sgen} is Bekenstein's `generalised entropy' \cite{Bekenstein:1974ax}, where $\sigma$ are taken to be cross-sections of event horizons. The generalised entropy is the sum of the geometric entropy and the entropy of the fields and, as a total entropy, it should satisfy a second law despite the shrinking of the area term due to Hawking evaporation.

Black hole evaporation takes place in the semiclassical approximation $G_{\mu\nu}=8\pi G\,\langle T_{\mu\nu}\rangle$. This approximation neglects the quantum fluctuations of the graviton, which is justified when the number of matter fields is large \cite{Wald:1984rg}.

When graviton fluctuations are important, one needs to define  $\left.S_{\textrm{bulk}}\right|_{\sigma}$ for the graviton in eq.~\eqref{eq:Sgen}. While this has not yet been achieved, it has been argued that  diffeomorphism invariance implies that the only surfaces $\sigma$ across which the entanglement of the graviton can be discussed are surfaces of extremal area \cite{Jafferis:2015del,  Camps:2018wjf}.

In other words, while one can consider any subregions of an ordinary system of fields, diff-invariance implies that the parts of the graviton are bounded by surfaces of extremal area. Below we explain why this is the case.

 \vspace{.8cm}

Any particular surface $\sigma$ in a given manifold ${\cal M}$ exists independently of coordinates. However, the only practical way to describe generic surfaces is through embedding maps $X^\mu(\sigma^i)$ -- where $\sigma^i$ span $\sigma$.  This makes explicit use of coordinates  $X^\mu$ in ${\cal M}$.

The metric $g$ of a pseudo-Riemannian manifold can be used to construct special families of coordinates, $X_g^\mu$. These correspond to gauge fixings of $g$. One example is the Fefferman-Graham gauge in asymptotically AdS geometries, that defines a radial coordinate $r$ via geodesics perpendicular to the conformal boundary. Embedding maps in Fefferman-Graham coordinates $r(\sigma^i)$ are indeed a very convenient way to characterise, e.g., Ryu-Takayanagi surfaces.

There are many sensible ways to gauge fix a metric, and hence to erect metric-related coordinate systems: $\{X^\mu_g, \tilde{X}^\mu_g,\cdots\}$. In a specific geometry $\bar{g}$ on ${\cal M}$, some of these distinctly-defined gauge fixings will actually describe the same coordinate system: $X^\mu_{\bar{g}}=\tilde{X}^\mu_{\bar{g}}$.


A surface of interest $\sigma_{\bar{g}}$ in a particular background geometry\footnote{From now on we slightly abuse notation and denote a geometry by $g$ rather than $({\cal M}, g)$.} $\bar{g}$ can be described with its embedding map in any coordinates. Obviously, the embedding maps are identical in the sets of coordinates that coincide in $\bar{g}$,  $X^\mu_{\bar{g}}(\sigma^i)=\tilde{X}^\mu_{\bar{g}}(\sigma^i)$. In fact, the class of equivalence of coordinates in $\bar{g}$ that describe $\sigma_{\bar{g}}$ with the same embedding map is much larger; any two coordinate systems that coincide on the surface $\sigma_{\bar{g}}$, but which may differ away from it, belong to the same equivalence class.

Consider now a surface across a family of background geometries $\{\bar{g}, g_2,g_3,\cdots\}$. That is, a set of pseudo-Riemannian manifolds each with one surface of interest, that we identify as ``the same surface across the set.''

Such surfaces across geometries $\sigma_g$ are necessary ingredients for any discussion of the quasilocal parts of the gravitational field; If we want to talk about which metric degrees of freedom have been excited on either side of a surface upon changing the geometry, we need to locate the surface of interest as the geometry changes.

While the discussion so far is classical, our main motivation is discussing quantum effects. We expect that, at least semiclassically, quantum states of the gravitational field $\psi$ can be characterised by Wigner functions\footnote{Wigner functions are phase space representations of quantum states. The Wigner function of a particle on a line, $W_\psi(x,p)$, has as position and momentum marginals the respective probability densities of the state $|\psi\rangle$ it represents: $|\langle x|\psi\rangle|^2=\int W_\psi(x,p)\,dp$, and $x\leftrightarrow p$.} over the phase space of General Relativity.  Phase space is the space of solutions of the equations of motion \cite{Crnkovic:1986ex}, and so in this description states spread over sets of solutions to the Einstein equations.

Given one such state $\psi$, peaked around $\bar{g}$ -- the classical limit of $\psi$ --, we want to discuss ``the entanglement of the graviton in $\psi$ across a surface $\sigma_{\bar{g}}$ in $\bar{g}$'' so as to define the second term in eq.~\eqref{eq:Sgen}. Our first task, then, is to extend $\sigma_{\bar{g}}$ in $\bar{g}$ to $\sigma_g$ in the family of geometries that make $\psi$.

Generically, the equivalence classes of embedding maps of $\sigma_g$ in the reference background geometry $\bar{g}$ split under changing the metric. That is, identical embedding maps in $\bar{g}$, $X^\mu_{\bar{g}}(\sigma^i)=\tilde{X}^\mu_{\bar{g}}(\sigma^i)$, result in different surfaces in other background geometries: $X^\mu_{g}(\sigma^i)\neq\tilde{X}^\mu_{g}(\sigma^i)$ as functions of $g$. For all practical purposes we can assume that the equivalence of coordinate systems in $\bar{g}$ is completely broken across generic families of geometries.

Extending a surface $\sigma_{\bar{g}}$ in $\bar{g}$ to $\sigma_g$ across the set of geometries via a particular embedding $X_g^\mu(\sigma^i)$ breaks the symmetry between equivalent coordinate systems in $\bar{g}$.  In other words, the description of a surface $\sigma_g$ across a set of background geometries via an embedding map $X^\mu_g(\sigma^i)$ is gauge-dependent. None of these coordinate grids $X_g^\mu$ is a natural structure in General Relativity. Only pseudo-Riemannian manifolds are available, not preferred sets of coordinates in them. The question of extending the definition of a generic $\sigma_{\bar{g}}$ in $\bar{g}$ to $\sigma_g$ in a set of $g$s is not well-posed.

There is, however, a special type of surfaces in $\bar{g}$ that extends naturally across families of geometries without the need of coordinates: the surfaces that are defined by their local geometry. The most important type of such surfaces for us have extremal area, i.e., zero trace of the extrinsic curvature:
\beq
K^\mu=0\,.
\label{eq:MinSurf}\eeq
It is straightforward to extend the definition of one such surface in $\bar{g}$ to a family of geometries without using any coordinates: in each of the background geometries $\sigma_g$ is an extremal surface subject to the same boundary conditions as in $\bar{g}$ -- e.g., being anchored at a specific location in a conformal boundary.

The description of surfaces via eq.~\eqref{eq:MinSurf} is coordinate invariant in the same way Einstein equations are: both equations describe local geometric facts.

Embedding maps relying on gauge fixings $X_g^\mu(\sigma^i)$  do have a geometric quality, but in contrast to eq.~\eqref{eq:MinSurf} they are generically non-local. For example, in Fefferman-Graham gauge, the coordinate $r$ is defined in relation to the conformal boundary of asymptotically AdS geometries. By contrast, General Relativity is both geometric and local.

The condition of having a local geometric definition would seem to allow many types of surfaces besides extremal-area ones. Could one consider, e.g., surfaces that extremise a curvature invariant other than the area?

This question can be asked precisely in the phase space formalism of General Relativity, and the answer is `no' \cite{Jafferis:2015del,  Camps:2018wjf}; In this formalism, diff-invariance requires that diffemorphisms annihilate the symplectic form, and translations of codimension-2 surfaces in General Relativity do so only for surfaces of extremal area.

Moreover, any functional other than the area would have to come accompanied by a lengthscale other than $1/G$ for it to be dimensionless. Such lengthscales are not justified in the semiclassical approximation.

It is however desirable to understand directly why more general geometrically defined surfaces are excluded; one speculation is that other extremisation problems are not  well-posed, and so it is impossible to find the desired surfaces across general enough families of geometries.

 \vspace{.8cm}

We have argued that the quasilocal parts of the gravitational field are bounded by surfaces that are geometrically and locally defined, i.e., that can be described without coordinates.

In General Relativity, the only acceptable such surfaces have extremal area. This implies, for example, that the entanglement of the graviton can only be discussed across extremal surfaces. This excludes the general cross-sections of event horizons of the generalised second law of thermodynamics. It does not preclude, however, a tentative generalised second law across the extremal-area HRT surfaces that can be associated to coarse-grained geometries outside `holographic screens' \cite{Engelhardt:2017aux}.

\subsection*{Acknowledgements}
Work funded by the `It from Qubit' Simons collaboration. I am grateful to Pablo Bueno, Shahar Hadar, and Aitor Lewkowycz for comments on the version submitted to the essay competition.

\end{document}